%% file: main.tex
\documentclass[journal]{IEEEtran}

\IEEEoverridecommandlockouts
\ifCLASSOPTIONcompsoc
  \usepackage[nocompress]{cite}
\else
  \usepackage{cite}
\fi

\ifCLASSINFOpdf
\else
\fi

\usepackage{indentfirst}
\usepackage{algorithm}
\usepackage{graphicx}
\usepackage{subfigure}
\usepackage{amsmath}
\usepackage{float}
\usepackage{booktabs}
\usepackage{epstopdf}
\usepackage{color}
\usepackage{subfigure}
\usepackage{url}
\usepackage{microtype}
\usepackage{tabularx}
\usepackage{algpseudocode}
\usepackage{amssymb}
\usepackage{color}

\newcommand{\etal}{{\it et al.~}}

\usepackage{tikz}

\usepackage{enumitem}

\newcommand{\bm}[1]{\mbox{\boldmath{$#1$}}}

\algdef{SE}[SUBALG]{Indent}{EndIndent}{}{\algorithmicend\ }%
\algtext*{Indent}
\algtext*{EndIndent}

\newcommand{\ra}[1]{\textcolor{black}{#1}}
\newcommand{\rb}[1]{\textcolor{black}{#1}}

\allowdisplaybreaks

\begin{document}

\title{\rb{PA-Cache: Evolving Learning-Based Popularity-Aware Content Caching\\ in Edge Networks}}

\author{Qilin Fan, ~\IEEEmembership{Member,~IEEE,}
Xiuhua Li, ~\IEEEmembership{Member,~IEEE,}
Jian Li, ~\IEEEmembership{Member,~IEEE,}\\
Qiang He, ~\IEEEmembership{Senior Member,~IEEE,}
Kai Wang, ~\IEEEmembership{Member,~IEEE,}
and Junhao Wen
\thanks{This work is supported in part by National NSFC under Grant No. 61902044 and 61672117, National Key R \& D Program of China under Grant No. 2018YFF0214700 and 2018YFB2100100, Chongqing Research Program of Basic Research and Frontier Technology under Grant No. cstc2019jcyj-msxmX0589 and cstc2018jcyjAX0340, Key Research Program of Chongqing under Grant No. CSTC2017jcyjBX0025 and CSTC2019jscx-zdztzxX0031, and the Fundamental Research Funds for the Central Universities under Grant No. 2018CDXYRJ0030 and 2020CDJQY-A022.
 \emph{(Corresponding author: Xiuhua Li.)}}%
\thanks{Q. Fan, X. Li and J. Wen are with Key Laboratory of Dependable Service Computing in Cyber Physical Society of Ministry of Education, Chongqing University, Chongqing 401331, China, with State Key Laboratory of Power Transmission Equipment and System Security and New Technology, Chongqing University, Chongqing 401331, China, and with the School of Big Data \& Software Engineering, Chongqing University, Chongqing 401331, China (e-mail: fanqilin@cqu.edu.cn; lixiuhua1988@gmail.com;  jhwen@cqu.edu.cn).}
\thanks{J. Li is with the Department of Electrical and Computer Engineering, Binghamton University, State University of New York, Binghamton, NY 13902, USA (e-mail: lij@binghamton.edu).}
\thanks{Q. He is with School of Software and Electrical Engineering, Swinburne University of Technology, Melbourne, 3122, Australia (e-mail: qhe@swin.edu.au).}
\thanks{K. Wang is with School of Computer Science and Technology, Harbin Institute of Technology, Weihai 264209, China, and with Research Institute of Cyberspace Security, Harbin Institute of Technology, Weihai 264209, China (e-mail: dr.wangkai@hit.edu.cn).}
}

\maketitle

\input{abstract}


\input{introduction}

\input{related-work}

\input{system}

\input{problem}
\input{design}

\input{evaluation}

\input{conclusion}

\ifCLASSOPTIONcaptionsoff
  \newpage
\fi

\bibliographystyle{IEEEtran}
\bibliography{main}

\end{document}

%% file: abstract.tex
\begin{abstract}

As ubiquitous and personalized services are growing boomingly, an increasingly large amount of traffic is generated over the network by massive mobile devices. As a result, content caching is gradually extending to network edges to provide low-latency services, improve quality of service, and reduce redundant data traffic. Compared to the conventional content delivery networks, caches in edge networks with smaller sizes usually have to accommodate more bursty requests. In this paper, we propose an evolving learning-based content caching policy, named PA-Cache in edge networks. It adaptively learns time-varying content popularity and determines which contents should be replaced when the cache is full. Unlike conventional deep neural networks (DNNs), which learn a fine-tuned but possibly outdated or biased prediction model using the entire training dataset with high computational complexity, PA-Cache weighs a large set of content features and trains the multi-layer recurrent neural network from shallow to deeper when more requests arrive over time. 
We extensively evaluate the performance of our proposed PA-Cache on real-world traces from a large online video-on-demand service provider. \rb{The results show that PA-Cache outperforms existing popular caching algorithms and approximates the optimal algorithm with only a 3.8\% performance gap when the cache percentage is 1.0\%}. PA-Cache also significantly reduces the computational cost compared to conventional DNN-based approaches.

\end{abstract}

%% file: introduction.tex
\section{Introduction}
\label{sec:introduction}

Today's Internet has seen an explosion of mobile data traffic in data-consuming application services such as mobile video services requested from a wide variety of mobile devices \cite{a-cooperative-green-content}.  According to the Cisco VNI report \cite{idc-report}, the overall average mobile traffic will increase at a compound annual growth rate of 46\% between 2017 and 2022, and it is estimated that video traffic will account for about 80\% of the global mobile data traffic. To enable ubiquitous mobile video services, \rb{cloud computing} has been widely acknowledged and deployed as it can provide adequate computing and storage resources. However, with the sky-rocketing network traffic load and more stringent requirements of users, cloud-based mobile video services are facing new challenges such as large transmission latency and \rb{limited bandwidth resources}.

To tackle the above challenges, edge computing has emerged as a new and evolving paradigm to accommodate future mobile video services, which provides computing and caching services close to \rb{end-users} at the network edges \cite{computation-offloading-with-multiple, enabling-collaborative-edge-computing}. In particular, the deployment of edge caching can  minimize the redundant data traffic, thereby leading to \rb{a} significant reduction in service latency and elimination in bandwidth wastage.

Edge caching strategy attempts to learn the pattern of content requests in some fashions \cite{online-collaborative-data-caching}, ensuring the availability of contents as high as possible in the cache nodes (\rb{e.g.,} base stations (BSs) and edge routers). 
Generally, the requested content is searched for in a cache node. If it is unavailable, a miss occurs, and the requested content is fetched from an upstream server (typically with higher latency and more expensive transmission cost). The content is then stored in the cache node and finally transmitted to the user. 
Besides, compared to the total size of contents, the capacity of cache nodes is usually limited and much smaller. If the new content is required to be cached when the cache is full, several cached contents may have to be evicted. Therefore, caching algorithms can also be described by the employed eviction strategy. 
When caching contents at network edges, user requests for the same content will be locally served. This can effectively reduce the redundant data traffic and greatly improve the quality of service \cite{zafari2019optimal}. 

Compared to the conventional content delivery networks (CDNs), edge caching has its unique characteristics \cite{intelligent-video-caching-at, cost-effective-app-data}: 
(i) \emph{Limited resources.} The cloud usually has a large number of diverse resources. However, the edge cache with limited computing and storage resources enables only a small fraction of contents to be cached and low-complexity tasks to be executed;
and (ii) \emph{Bursty requests.} The requests from edge networks usually vary a lot over time.  

\rb{\rb{Nowadays, most caching system still utilize recency-based \cite{analyzing-the-performance-of}, frequency-based \cite{high-performance-cache-replacement}, size-based \cite{caching-proxies-limitations-and}, or combinations of them.}
The limitation is that they might work well for some access patterns but poorly for others. Recently, learning-based caching algorithms have been proposed either to determine which contents should be evicted when the cache is full or decide whether or not to admit a content upon a request by learning content popularity. However, the content requests of edge networks are time-varying and bursty. \rb{On the one hand, it is difficult for shallow machine learning models to capture complex patterns}. On the other hand, by using the entire training dataset, conventional deep neural networks (DNNs) would learn a fine-tuned but possibly outdated or biased prediction model with high computation complexity, \rb{making is difficult to support} the application at edge caches with limited computing capability.
}


In this paper, we propose a novel popularity-aware content caching policy, namely PA-Cache, in edge networks. Different from previous approaches, PA-Cache weighs a large set of content features to learn the time-varying content popularity adaptively. Furthermore, to overcome the \rb{high} computational cost of conventional DNN, PA-Cache takes advantage of \rb{a} shallow network with fast convergence at the beginning, and then the powerful representation of DNN when more requests arrive over time. In this way, PA-Cache achieves high scalability and high computation efficiency. The contributions of this paper can be summarized as follows:

 \begin{itemize}
\item We investigate the issue of popularity-aware content caching and design modules and operations of popularity-aware cache nodes in edge networks. We apply \rb{a} learning-based approach to tackle this problem. 
\item We amend the multi-layer recurrent neural network (RNN) architecture by attaching every hidden layer representation to an output regression to predict the temporal content popularity. We utilize a hedge strategy, which enables adaptive training of DNN in an evolving setting. The popularity-aware cache node makes appropriate cache replacement decisions to maximize the long-term cache hit rate based on the estimated popularity.
\item We conduct extensive experiments on a real-world dataset derived from iQiYi, which is the largest online video-on-demand (VoD) service provider in China. Trace-driven evaluation results demonstrate the effectiveness and superiority of PA-Cache over several candidate algorithms.
\end{itemize}

The rest of this paper is organized as follows. The related work is briefly introduced in Section \ref{related-work}. Section \ref{system} provides a system overview. Section \ref{problem} gives a formal description of the cache replacement problem. 
Section \ref{design} presents \rb{the design of PA-Cache algorithm in detail}.
In Section \ref{evaluation}, the performance of our proposed algorithm by trace-driven experiments is evaluated. Section \ref{conclusion} concludes this paper.

%% file: related-work.tex
\section{Related Work} 
\label{related-work}

Content caching algorithms have been studied for many decades. Existing work can be divided into the following two main branches.

\subsection{\rb{Rule-Based Algorithms}}
\rb{\rb{The first branch is named rule-based algorithms. These cache eviction algorithms rely on one of the most widely used features (i.e., recency, frequency and size) or their combinations.
LRU-K \cite{the-LRU-K-page} is a combination of least recently user (LRU) \cite{analyzing-the-performance-of} and least frequently used (LFU) \cite{high-performance-cache-replacement}, in which \rb{the cache} remembers the time of last K occurrences instead of \rb{the} last occurrence for each content.}
ARC \cite{arc-a-self-tuning} adaptively divided the cache space into two segments and maintains contents that have been referenced exactly once and at least twice, respectively.
S4LRU \cite{an-analysis-of-facebook} \rb{partitioned} the cache into four lists, and each list \rb{was an} LRU queue.  Vietri \etal \cite{driving-cache-replacement-with} proposed \rb{a} LeCaR algorithm which \rb{managed} two histories (i.e., recency and frequency) of metadata for each cache entry, and their weights \rb{were} adaptively updated by regret minimization. 
\rb{
Bahn \etal \cite{efficient-replacement-of-nonuniform} evaluated the content based
on its past accesses to estimate the likelihood of re-access, defined a  least-unified value metric and normalizes it by the cost per unit size. 
Berger \etal \cite{practical-bounds-on-optimal} formulated the caching as a min-cost flow problem when considering variable content sizes.
}
}

To evaluate the performance of classic cache eviction algorithms, Martina \etal \cite{a-unified-approach-to} proposed a unified and flexible approach by extending and generalizing a fairly decoupling principle (the \rb{so-called} Che's approximation \cite{hierarchical-web-caching-systems}). These algorithms follow heuristic rules and are easy to implement. However, most of the analysis was derived under \rb{an} independent reference model, assuming that content popularity follows a fixed Zipf law \cite{web-caching-and-zipf}, \rb{i.e., $p_{i} \propto i^{-\alpha}, \alpha>0$, where $p_{i}$ refers to the request probability of $i$-th most popular content.} The performance of a cache eviction algorithm under synthetic data traces is found to be quite different from that under real data traces \cite{accurate-learning-or-fast}.
Therefore, these algorithms might hardly accommodate the dynamic content access pattern in edge networks.

\subsection{\rb{Machine Learning-Based Algorithms}}
The second branch relies on machine learning algorithms to optimize the content caching strategy. This branch is further subdivided into two categories.

The first category investigates the use of ``model-free'' reinforcement learning (RL) \rb{in which} the system starts without any prior assumption about the traffic pattern to converge towards the optimal caching strategy \cite{a-deep-reinforcement-learning,rl-cache-learning-based,deep-reinforcement-learning-for,federated-deep-reinforcement-learning,na-caching-an-adaptive,caca-learning-based-content,yan2020rl}. Such system learns to make decisions from experience through interactions with the environment, in which good actions are enhanced via a reward function. 
Kirilin \etal \cite{rl-cache-learning-based}  trained a feedforward neural network (FNN) using a Monte Carlo method, which \rb{computed} the admission probability to maximize the cache hit rate. 
Sadeghi \etal \cite{deep-reinforcement-learning-for} developed a two-timescale deep Q-network approach in a hierarchical cache network to learn an online caching policy.
Wang \etal \cite{federated-deep-reinforcement-learning} formulated the content caching procedure as a Markov decision process and utilized double deep Q-network which \rb{stabilized} the learning in discrete huge spaces for training the caching process.
\rb{Fan \etal \cite{na-caching-an-adaptive} leveraged the beneﬁts of the RNN as well as the Deep Q-Network to maximize the cache efﬁciency by jointly learning request features, caching space dynamics, and making decisions.}
Guan \etal \cite{caca-learning-based-content} designed a content-aware cache admission strategy. It consists of a tree-structured learning model for mining the critical features and a UCB-tree algorithm for making cache admission decisions dynamically. 
\rb{Yan \etal \cite{yan2020rl} leveraged RL and Belady to learn content popularity for both content admission and content eviction.}  
However, RL-based algorithms require a tremendous number of learning samples and suffer from large-delayed rewards (cache hits). This can result in slow reaction times in highly dynamic environments \cite{towards-lightweight-and-robust}. Furthermore, as RL-based algorithms are quite sensitive to hyperparameters and random seeds, it is difficult to
configure and maintain these algorithms \cite{deep-reinforcement-learning-that}. 

The second category utilizes supervised learning, which learns key attribute features in requests and predicts the popularity or arrival times of contents to facilitate efficient caching \cite{feedforward-neural-networks-for, toward-smart-and-cooperative, learning-relaxed-belady-for,iotcache-toward-data-driven}. 
\rb{Fedchenko \etal \cite{feedforward-neural-networks-for} employed FNN to predict content popularity and made caching decisions  accordingly.}
Pang \etal \cite{toward-smart-and-cooperative} utilized deep long short-term memory (LSTM) to calculate the probability for each content to arrive.
Song \etal \cite{learning-relaxed-belady-for} explored gradient boosting machine model to predict log (time-to-next-request). 
Chen \etal \cite{iotcache-toward-data-driven} designed a DNN-based popularity evolving model to estimate the near future popularity of IoT data. However, \rb{on the one hand}, shallow machine learning models would be difficult to learn complex patterns. On the other hand, by using the entire training dataset, conventional DNNs would learn a fine-tuned but possibly outdated or biased prediction model with high computation complexity. Only a few papers \cite{popularity-driven-content-caching,context-aware-proactive-content} proposed to use online learning-based algorithms as an adaptation method to predict content popularity by clustering the contents according to their similarities of access patterns. However, the selection of hand-crafted features has a significant impact on its performance gain. Moreover, the prediction model, which is essentially based on the sample average approximation method, is still shallow.

%% file: system.tex
\section{System Overview} 
\label{system}

\rb{In this section, we firstly introduce the cache node structure in the considered edge network. Then, we describe the main operations of the edge node.}

\subsection{Cache Node Structure}

\begin{figure}
\centering
\includegraphics[width=0.95\linewidth]{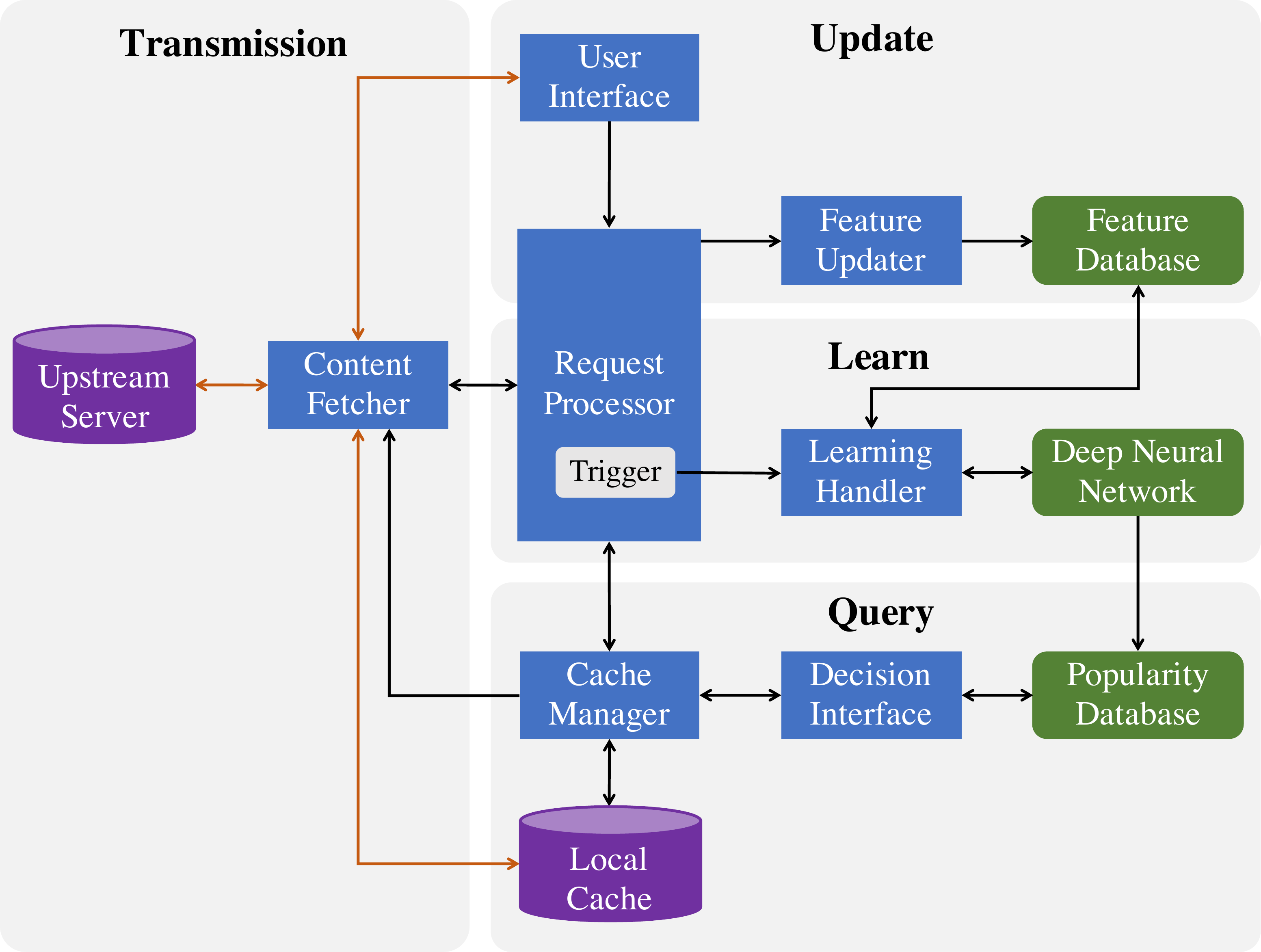}
\caption{Modules and operations of a single popularity-aware cache node in an edge network.}
\label{fig:structure}
\end{figure}

\rb{The edge caching service is implemented and provided collectively by multiple cache nodes. Fig. \ref{fig:structure} illustrates the modules of a considered popularity-aware cache node in an edge network. \ra{It provides both storage and computation capacity for caching service. As high-performance CPUs and GPUs are enabled (e.g., NVIDIA Jetson TX21\footnote{https://developer.nvidia.com/embedded-computing}), deep learning tasks can be processed in the cache node.} Typically, a cache node contains basic models: \textit{Request Processor}, \textit{Cache Manager}, \textit{User Interface}, \textit{Content Fetcher}, \textit{Local Cache}. A popularity-aware cache node also integrates with \textit{Feature Updater}, \textit{Learning Handler}, \textit{Decision Interface}, two databases (i.e., \textit{Feature Database}, \textit{Popularity Database}), as well as a \textit{Deep Neural Network} to enable popularity awareness.
}

1) The \textit{Feature Updater} module updates features from raw data of content requests, including contextual features (e.g., the number of video requests) and semantic features (e.g., video type) of content, and storing them in the \textit{Feature Database} with a unified format.

2) The \textit{Learning Handler} module trains a \textit{Deep Neural Network} to predict content popularity in the next time step from the data in \textit{Feature Database}.

3) The \textit{Decision Interface} module determines which cached content should be evicted when a cache miss is declared.

\subsection{Operations}

The main operations of a popularity-aware cache node consist of three procedures:

1) \textbf{Update.} The \textit{Request Processor} receives a request and initiates an update operation based on the metadata of the received request. Then, the \textit{Feature Updater} keeps track of the raw data, extracts the latest features, and writes them to the \textit{Feature Database}.

2) \textbf{Learn.} The \textit{Request Processor} periodically triggers a \textit{Learning Handler} to \rb{learn a prediction model every $\phi$ hours}. Parameter $\phi$ requires a proper selection for balancing the high accuracy against the low computational cost for the prediction model. The \textit{Learning Handler} extracts the raw features and ground-truth popularity from the \textit{Feature Database}. Then, it takes the normalized features together with the ground-truth popularity as the input to train a \textit{Deep Neural Network} in an evolving manner. The predicted content popularity calculated by \textit{Deep Neural Network} is recorded in the \textit{Popularity Database}.

3) \textbf{Query.} The \textit{Cache Manager} examines if the requested content is in the \textit{Local Cache}. When it is locally available, the \textit{Content Fetcher} fetches the content from the \textit{Local Cache} and serves the user. Otherwise, the \textit{Cache Manager} sends a query to the \textit{Decision Interface} that determines which cached content should be evicted according to \textit{Popularity Database} and \rb{get its response}. The \textit{Cache Manager} evicts the least popular content and notifies the \textit{Content Fetcher} to fetch the requested content from the \textit{Upstream Server}, store it in the \textit{Local Cache} and serve the user.

%% file: problem.tex
\section{Problem Statement}
\label{problem}

\rb{In this paper, let $\mathcal{C} = \left \{1,2, \dots, C\right\}$ denote a given set of contents that are distributed to multiple end-users associated with the cache node.  
\rb{Since many replication strategies in the VoD system fragment the contents into equally sized chunks \cite{ low-complexity-content-replication}, we assume that all contents are unit-sized. It reduces the complexities and inefficiencies of continually allocating and de-allocating storage space to contents with non-uniform sizes. However, our work could also be extended to the arbitrary sizes considering popularity per unit size at the eviction phase like previous works \cite{efficient-replacement-of-nonuniform, content-and-network-aware}.} Let the cache capacity be $s$, that is, the cache node can accommodate up to $s$ contents.
The sequence of requests for content is denoted as $\mathcal{K} = \left \{1,2, \dots, K\right\}$.
Each request $k\in\mathcal{K}$ in this sequence is composed of three elements: the requested content $c_k$, the timestamp $t_k$ and the $d$-dimensional feature vector of the requested content $\bm{f}_k$.
}

For each request $k$, we have to examine whether it can be served by the cache. 
For this  purpose, let $\bm{Z}_k = [Z_k^1,Z_k^2,\dots,Z_k^C]$ be the indication vector of cache status at time $t_k$, where $Z_k^c \in \{0, 1\}$ is an indicator of whether content $c$ is in the cache or not (i.e., $Z_k^c=1$ indicates that $c$ is available in the local cache and current request can be served, and 0 otherwise.)

When content $c_k$ is hit in the cache, the cache status vector stays the same: $\bm{Z}_{k+1} = \bm{Z}_k$. Otherwise, when content $c_k$ suffers from a miss, the cache node will retrieve it from a specific upstream server according to the traffic assignment criteria \cite{towards-optimal-request-mapping}. In this situation, the cache node will remove an old content $c^{evict}$ to make room for the new content $c_k$. Formally, the cache state transition can be modelled as follows:
\begin{equation}
Z_{k+1}^c =\left\{
\begin{array}{lcl}
0,       	&      &  c = c^{evict},\\
1,     		&      &  c = c_k, \\
Z_k^c,    &      &  \text{otherwise}.\\
\end{array} \right.
\end{equation}

Whenever a request $k$ arrives, a caching policy $\pi$ maps the current cache status  vector $\bm{Z}_k$, the requested content $c_k$ and the feature vector of the requested content $\bm{f}_k$ to the new cache status vector $\bm{Z}_{k+1}$. The cache status vector can be updated based on $\pi$ as follows:
\begin{equation}
\bm{Z}_{k+1} = \pi(\bm{Z}_k|c_k,\bm{f}_k).
\end{equation}

We use the term $H_{\pi}(K)$ as a cache hit rate metric to evaluate the efficiency of the caching system. It is defined as the ratio of requests that are served from the local cache to $K$ requests, which is given by:
\begin{equation}
H_{\pi}(K)= \frac{1}{K}\sum_{k} Z_k^{c_k}.
\end{equation}

Furthermore, $H_{\pi}$ is introduced to represent the long-term average cache hit rate when the number of requests goes to infinity over time by adopting caching policy $\pi$, which is written as:
\begin{equation}
H_{\pi} = \lim_{K\to\infty} H_{\pi}(K).
\end{equation}

\rb{In the considered edge network, our objective is to maximize $H_{\pi}$ based on the constraint of cache capacity, and then find a policy $\pi$ for generating a series of popularity-aware replacement actions. Consequently, the corresponding problem can be formulated as:}
\begin{align}
\max \limits_{\pi} \quad	&  	H_{\pi} \\
s.t. \quad	& \sum_{c} Z_k^c \leq s, \quad \forall k, \\
      \quad & Z_k^c \in \left\{0,1\right\}, \quad \forall k, \forall c.
\end{align}

%% file: design.tex
\section{PA-Cache Algorithm Design}
\label{design}

\rb{In this section, we present the PA-Cache, a popularity-aware content caching approach that makes appropriate cache replacement decisions to maximize the long-term cache hit rate based on the estimated popularity. We begin by introducing the basic idea of the optimal replacement policy. We then present a deep learning-based approach for predicting the content popularity. Finally, we describe how the PA-Cache makes online replacement decisions.}

\subsection{Basic Idea}
\rb{Van Roy \etal \cite{a-short-proof-of} presented a proof showing that the replacement policy, named MIN \cite {a-study-of-replacement} proposed by Belady is an optimal policy of the above problem.} The MIN policy replaces the content in the cache, which has the longest time to be visited next time. As the future information is required in advance, it is an idealistic algorithm which is unimplementable in a real system. However, Belady's MIN algorithm gives a performance upper bound for content caching algorithms.

\begin{table}[t]
\centering
\normalsize
\caption{Detailed Features}\label{tab:feature}
\begin{tabular}{|l|l|}
\hline
\textbf{Category}           & \textbf{Feature}                                                                                                                     \\ \hline
contextual feature & \begin{tabular}[c]{@{}l@{}}access times in previous time step \\ age = current time - publish time\end{tabular}    \\  \hline
semantic feature   & \begin{tabular}[c]{@{}l@{}}type\\ length\\ area\\ language\\ score\\ number of comments\\ director\\ performer\end{tabular} \\ \hline
\end{tabular}
\end{table}

It is a challenging task to imitate the MIN algorithm using learning-based approaches directly. To pick out the content whose next request time is farthest in the future, learning-based approaches are required to predict the next request time of all contents in the cache precisely. Besides, running a predictor for all contents in the cache upon each request will consume significant computing resources and time.

Therefore, in this paper,  we aim to approximate \rb{the} MIN algorithm by predicting content popularity within a specific time interval/step to address \rb{the} aforementioned issues. We propose a popularity-aware content caching algorithm, named PA-Cache, which consists of two phases: 
\textit{offline content popularity prediction} (in Section \ref{offline}) and \textit{online replacement decision} (in Section \ref{online}).

\subsection{Offline Content Popularity Prediction} 
\label{offline}

As motivated earlier, we consider an evolving prediction task. The time period is partitioned into consecutive time steps indexed by $t=1,\dots,T$. Our goal of evolving deep learning is to learn a function $F : \mathbb{R}^{m \times d} \rightarrow \mathbb{R}^m$ based on a sequence of training samples $\mathcal{D} = \{(\bm{x}_1, \bm{y}_1), \dots, (\bm{x}_t, \bm{y}_t),\dots, (\bm{x}_T, \bm{y}_T)\}$, that arrive sequentially, where $\bm{x}_t \in \mathbb{R}^{m \times d}$ represents the input features at time step $t$, $m$ is the number of instances, and $d$ is the feature dimension. $(\bm{x}_t^i, y_t^i)$ is the $i^{th}$ sample at time step $t$. The corresponding ground-truth content popularity is denoted as $\bm{y}_t \in \mathbb{R}^m$ while the predicted content popularity is denoted as $\hat{\bm{y}}_t \in \mathbb{R}^m$. The performance of learnt model is evaluated in terms of the cumulative prediction error of $m$ mini-batch instances.

\subsubsection{Feature Selection}

We consider two main types of features: the contextual features that would be time-varying; and the semantic features that are invariant with time. The detailed features are given in Table \ref{tab:feature}. As the categorical features such as type and language are non-numeric features, which would not have a natural rank-ordering, we transform them into one-hot coding vectors, which is widely utilized in word embedding area \cite{distributed-representations-of-words}.
Furthermore, as the numeric features (e.g., access times in previous time step, length) might have a wide range of values, their values are normalized into the range of $[0,1]$.

\subsubsection{Evolving Deep Learning Model}

\begin{figure}[t]
\centering
\includegraphics[width=0.95\linewidth]{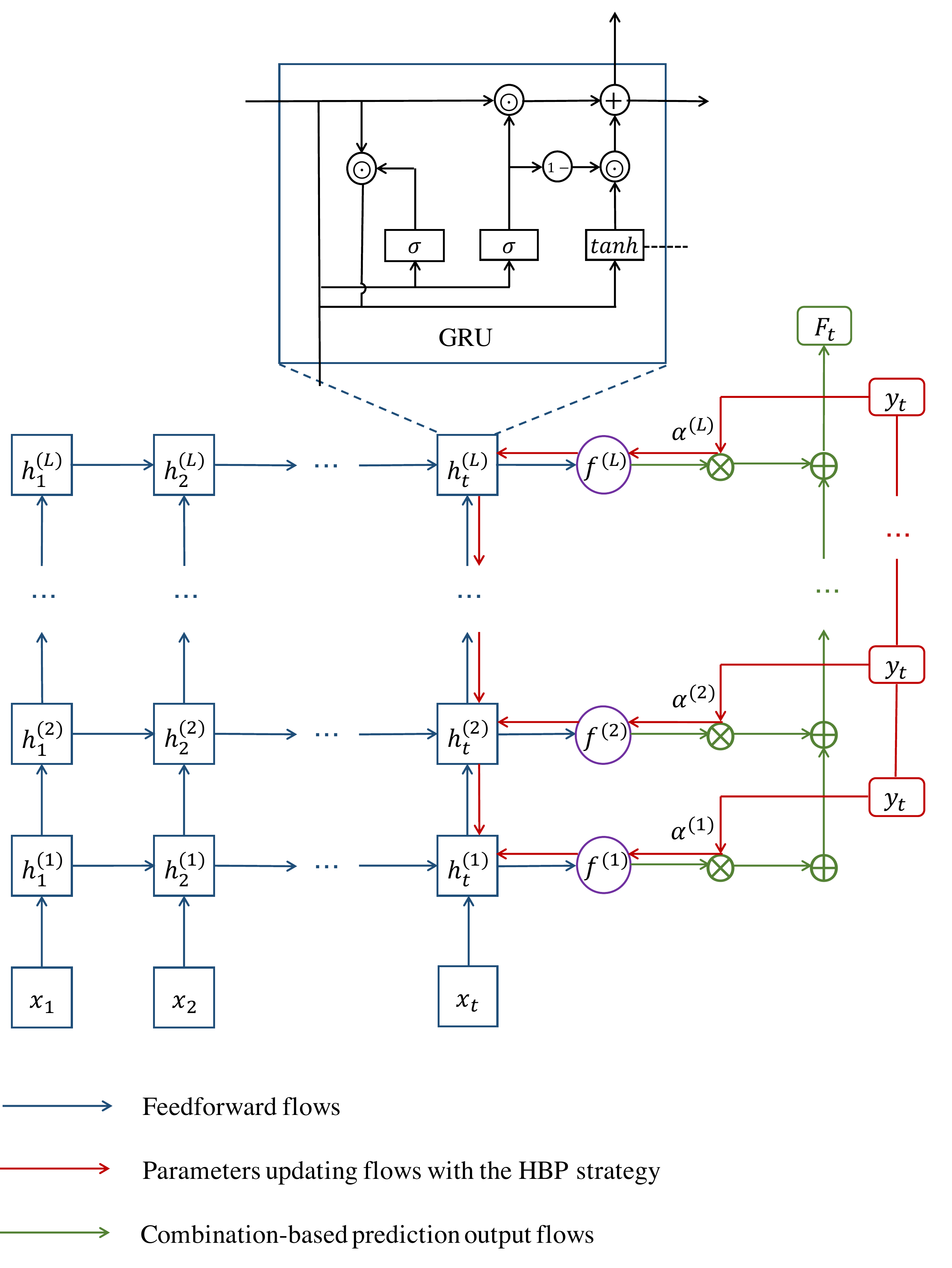}
\caption{Evolving deep learning framework using HBP.}
\label{fig:model}
\end{figure}

Given an input $\bm{x_t}$, the content popularity prediction task can be conducted by a conventional multi-layer recurrent neural network (RNN) with $L$ hidden layers. However, using such a model for evolving learning faces several challenges \cite{online-deep-learning-learning}: (i) {\textit{Model selection}. The network depth must be fixed in advance and cannot be changed.  However, selecting an appropriate depth is a daunting task, especially in an evolving environment.  For a small number of instances, a shallow network would be preferred as it converges quickly, while for a large number of instances, a deep network would be better at generalization and accurate prediction; (ii) {\textit{Convergence}. Critical issues of deep architecture consist of exploding or vanishing gradients, saddle point and diminishing feature reuse. These issues will result in slow or unstable convergence and be further exaggerated in the evolving setting.

To address these issues, we amend the multi-layer RNN architecture by attaching every hidden layer representation to an output regression for evolving learning through hedge backpropagation (HBP) \cite{online-deep-learning-learning}.  HBP automatically determines how and when to accommodate network depth in an evolving manner. An evolving deep learning framework using HBP is depicted in Fig. \ref{fig:model}.

In this paper, we consider a multi-layer RNN with $L$ hidden layers, where $L$ denotes the maximum depth of the network.  
Feature vectors are fed into a gated recurrent unit (GRU) layer, which is capable of capturing the long-term dependencies in sequential data and learning the time-varying patterns of user requests.
The standard GRU architecture is based on two multiplicative gate units. The prediction function $F(\bm{x}_t) $ for the proposed evolving DNN is given by: 
\begin{align}
F(\bm{x}_t) 		  &= \sum_{l=1}^L \alpha^{(l)}f^{(l)}(\bm{x}_t), \quad \text{where} \\
f^{(l)}(\bm{x}_t)  &= \bm{\Theta}^{(l)}\bm{h}_{t}	^{(l)},	\notag \\
\bm{r}_{t}^{(l)}		  &= sigm(\bm{W}_r^{(l)} \bm{h}_{t}^{(l-1)} + \bm{U}_r^{(l)}\bm{h}_{t-1}^{(l)} + \bm{b}_r^{(l)}),\notag \\
\bm{z}_{t}^{(l)}		  &=sigm(\bm{W}_z^{(l)} \bm{h}_{t}^{(l-1)} + \bm{U}_z^{(l)}\bm{h}_{t-1}^{(l)} + \bm{b}_z^{(l)}), \notag\\
\bm{h}_{t}^{(l)}              &= \bm{z}_{t}^{(l)} \odot \bm{h}_{t-1}^{(l)} + (\bm{1} - \bm{z}_{t}^{(l)})\odot \tilde{\bm{h}}_{t}^{(l)},\notag \\
\tilde{\bm{h}}_{t}^{(l)} 	&=  tanh(\bm{W}_h^{(l)}\bm{h}_t^{(l-1)}  + \bm{U}_h^{(l)}  (\bm{r}_t^{(l)}  \odot \bm{h}_{t-1}^{(l)}  )+ \bm{b}_h^{(l)}  ), \notag\\
\bm{h}_t^{(0)}                   &= \bm{x}_t,\notag
\end{align}
where $\bm{r}_t$, $\bm{z}_t$, $\bm{h}_t$, $\tilde{\bm{h}}_t$ are referred to as reset gate, update gate, hidden state vector and candidate hidden state vector respectively. The reset gate is used to specify how much previous knowledge should be ignored. The update gate helps the model to control the information that flows into memory. The candidate hidden state, also named intermediate memory unit, combines the knowledge from hidden states of \rb{the} previous time step and previous layer. It can facilitate the computation of subsequent hidden state.
The parameters for training are the matrices $\left \{\bm{W}_r, \bm{W}_z, \bm{W}_h\right \}$ (the feedforward connection weights), $\left \{\bm{U}_r, \bm{U}_z, \bm{U}_h\right \}$ (the recurrent weights) and the vectors $\left \{\bm{b}_r, \bm{b}_z, \bm{b}_h\right \}$ (the bias). $sigm(\cdot)$ is the sigmoid activation function, which limits $\bm{r}_t$ and $\bm{z}_t$ to take values ranging from 0 and 1. The element-wise tensor multiplication is denoted by $\odot$.

Different from the conventional DNN, in which only the hidden state vector of last layer $\bm{h}^{(L)}$ is used by the regression to calculate the final predicted value, we utilize a weighted combination of regressions learned based on the multiple hidden state vectors from $\bm{h}^{(1)}, \dots , \bm{h}^{(L)}$ in this paper.
Two sets of new parameters to be learnt are introduced, i.e., $\bm{\Theta}^{(l)}$ and $\alpha^{(l)}$. Each regression in intermediate layer $f^{(l)}(\bm{x}_t)$  is parameterized by $\bm{\Theta}^{(l)}$. The final prediction of this model is a linear weighted sum of regressions $f^{(1)}(\bm{x}_t), \dots , f^{(L)}(\bm{x}_t)$ , where the weight of each regression is denoted by $\alpha^{(l)}>0$. The loss function of this model is defined as: $\mathcal{L}(F(\bm{x}_t),\bm{y}_t) = \sum_{l=1}^L \alpha^{(l)} \mathcal{L}(f^{(l)}(\bm{x}_t),\bm{y}_t)$. During the evolving learning procedure, parameters $\alpha^{(l)}$, $\bm{\Theta}^{(l)}$, $\bm{W}^{(l)}$, $\bm{U}^{(l)}$, $\bm{b}^{(l)}$ are required to be learnt.

We employ the HBP strategy \cite{a-decision-theoretic-generalization} to learn $\alpha^{(l)}$. In the beginning, all weights $\alpha^{(l)}$ are uniformly split, i.e., $\alpha^{(l)} = \frac{1}{L}, l = 1,\dots,L$. At each iteration, the regression of layer $l$ makes a prediction $f^{(l)}(\bm{x}_t)$. The weight of the regression then is updated as follows:
\begin{equation}
\alpha^{(l)}_{t+1} \leftarrow \alpha^{(l)}_t \beta^{\min(\mathcal{L}(f^{(l)}(\bm{x}_t),\bm{y}_t), \kappa)},\label{equ:updatealpha}
\end{equation}
where $\beta \in (0, 1)$ is the discount factor, $\kappa$ is the threshold parameter for smoothing ``noisy" data. Thus, a regression's weight decays of a factor of $\beta^{\min(\mathcal{L}(f^{(l)}(\bm{x}_t),\bm{y}_t), \kappa)}$ in each iteration. At the end of each iteration, weights $\alpha$ are normalized to the interval $[0, 1]$, i.e., $ \sum_{l=1}^L \alpha^{(l)}=1$.

We adopt gradient descent methods to learn the parameters $\bm{\Theta}^{(l)}$ for all regressions, where the input to the $l^{th}$ regression is $\bm{h}^{(l)}$. It is identical with the update of the weights of the output layer in the conventional feedforward neural networks. This update is given by: 
\begin{align}
\bm{\Theta}^{(l)}_{t+1} &\leftarrow \bm{\Theta}^{(l)}_t - \eta\nabla_{\bm{\Theta}^{(l)}_t}\mathcal{L}(F(\bm{x}_t),\bm{y}_t)  \label{equ:updatetheta}\notag \\
&\leftarrow \bm{\Theta}^{(l)}_t - \eta\alpha^{(l)}\nabla_{\bm{\Theta}^{(l)}_t}\mathcal{L}(f^{(l)}(\bm{x}_t),\bm{y}_t).
\end{align}

Updating the feature representation parameters $\bm{W}^{(l)}, \bm{U}^{(l)}$, $\bm{b}^{(l)}$ is not trivial. Unlike the original backpropagation scheme, in which the error derivatives are backpropagated from the output layer to each hidden layer, the error derivatives in this paper are backpropagated from each regression $f^{(l)}(\bm{x}_t)$. Thus, by combining the gradient descent methods and the dynamic objective function $\mathcal{L}(F(\bm{x}_t),\bm{y}_t) = \sum_{l=1}^L \alpha^{(l)} \mathcal{L}(f^{(l)}(\bm{x}_t),\bm{y}_t)$, the update rule for $\bm{W}^{(l)}$, $\bm{U}^{(l)}$, $\bm{b}^{(l)}$ is given as follows:
\begin{align}
\bm{W}^{(l)}_{t+1} &\leftarrow \bm{W}^{(l)}_t - \eta\sum_{j=l}^{L}\alpha^{(j)}\nabla_{\bm{W}^{(l)}_t}\mathcal{L}(f^{(j)}(\bm{x}_t),\bm{y}_t), \label{equ:updatew}\\
\bm{U}^{(l)}_{t+1} &\leftarrow \bm{U}^{(l)}_t - \eta\sum_{j=l}^{L}\alpha^{(j)}\nabla_{\bm{U}^{(l)}_t}\mathcal{L}(f^{(j)}(\bm{x}_t), \bm{y}_t), \label{equ:updateu}\\
\bm{b}^{(l)}_{t+1} &\leftarrow \bm{b}^{(l)}_t - \eta\sum_{j=l}^{L}\alpha^{(j)}\nabla_{\bm{b}^{(l)}_t}\mathcal{L}(f^{(j)}(\bm{x}_t), \bm{y}_t), \label{equ:updateb}
\end{align}
where $\nabla_{\bm{W}^{(l)}_t}\mathcal{L}(f^{(j)}(\bm{x}_t), \bm{y}_t))$, $\nabla_{\bm{U}^{(l)}_t}\mathcal{L}(f^{(j)}(\bm{x}_t), \bm{y}_t))$, $\nabla_{\bm{b}^{(l)}_t}\mathcal{L}(f^{(j)}(\bm{x}_t), \bm{y}_t))$ are computed via backpropagation from error derivatives of $f^{(j)}(\bm{x}_t)$. The summation part in Eq. (\ref{equ:updatew})-(\ref{equ:updateb}) starts at $j = l$ as the shallower regression does not rely on the parameters $\bm{W}$, $\bm{U}$, $\bm{b}$ of deeper layers to make predictions.

Since shallower models are usually inclined to converge faster than deeper models \cite{fractalnet-ultra-deep-neural}, the weights of deeper regression might be diminished to a very small value by the HBP strategy. This will lead to a slow convergence in deeper regressions. Therefore, a smoothing parameter $\zeta \in (0, 1)$ is introduced to set a minimum weight for each regression. After the weight of the regressions are updated in each iteration according to Eq.(\ref{equ:updatealpha}), we further set the weights as $\alpha^{(l)} \leftarrow \max(\alpha^{(l)},\frac{\zeta}{L})$, where $\zeta$ guarantees that each regression will be selected with at least probability $\frac{\zeta}{L}$. This balances the trade-off between exploration (all regressions at every depth will affect the backpropagation update) and exploitation. 

\subsubsection{Loss Function}
Typically, DNN tracks the mean square error (MSE) as a metric when fitting the model. However, this metric is unable to make \rb{an} accurate prediction when applied to the data that follows the heavy-tailed distribution, which is common in VoD systems \cite{on-the-familiar-stranger}. It is because popular contents with high access times only account for a small fraction, but the arithmetic mean will be biased towards them. Hence, the overall performance of this metric would not be so satisfactory for the majority of contents. To eliminate this drawback, a mean relative squared error (MRSE) \cite{using-early-view-patterns} is employed here, which is written as:
\begin{equation}
\mathcal{L}(f^{(l)}(\bm{x}_t),\bm{y_t}) = \frac{1}{m}\sum_{i=1}^m(\frac{f^{(l)}(\bm{x}_t^i)}{y_t^i}-1)^2.
\label{equ:loss}
\end{equation}

Algorithm \ref{algorithm1} outlines the popularity prediction using evolving deep learning.

\begin{algorithm}[t]
	\caption{Popularity prediction using evolving deep learning.}\label{algorithm1}
	\hspace*{0.02in} {\bf Input:} Input features: $\bm{x}_t$; Revealed popularity: $\bm{y}_t$; Parameters for evolving DNN: $\left\{ \beta, \eta, \zeta, \kappa, \alpha_t, \bm{\Theta}_t, \bm{W}_t, \bm{U}_t, \bm{b}_t \right\} $ \\
	\hspace*{0.02in} {\bf Output:} Predicted popularity: $\hat{\bm{y}}_t$
	\begin{algorithmic}[1]
	 	\State $\hat{\bm{y}}_t = F(\bm{x}_t) =
\sum_{l=1}^L \alpha_t^{(l)}f^{(l)}(\bm{x}_t)$;
	 	\State Calculate $\mathcal{L}^{(l)}_t = \mathcal{L}(f^{(l)}(\bm{x}_t),\bm{y}_t),  \forall l=1,\dots,L$ by Eq. (\ref{equ:loss});
	 	\State Update $\bm{\Theta}^{(l)}_{t+1}, \bm{W}^{(l)}_{t+1}, \bm{U}^{(l)}_{t+1}, \bm{b}^{(l)}_{t+1}, \forall l=1,\dots,L$ by Eq. (\ref{equ:updatetheta}) - Eq. (\ref{equ:updateb}); 
	 	\State Update $\alpha^{(l)}_{t+1} = \alpha^{(l)}_t \beta^{ \min(\mathcal{L}^{(l)}_t, \kappa)}, \forall l=1,\dots,L$;
	 	\State Smoothing $\alpha^{(l)}_{t+1} = \max(\alpha^{(l)}_{t+1},\frac{\zeta}{L}), \forall l=1,\dots,L$;
		\State Normalize $\alpha^{(l)}_{t+1} =\frac{\alpha^{(l)}_{t+1}}{Z_{t+1}}$ where $Z_{t+1} =\sum_{l=1}^{L}\alpha^{(l)}_{t+1}$;
	\end{algorithmic}
\end{algorithm}

\subsection{Online Replacement Decision}
\label{online}
When a request $k$ of content $c_k$ arrives, we first extract its latest features of the requested content and write them into the \textit{Feature Database} module. 
Then, the requested content is searched for in the local cache. If it is available in the cache, the user is directly served by the content replica from the cache node. Otherwise, PA-Cache removes the least popular content $c^{evict}$ in the cache to leave space for the new content $c_k$.
\textit{Popularity Database} module provides the  estimated popularity which is updated by \textit{Deep Neural Network} periodically described in Section \ref{offline}. \rb{Afterward}, $c_k$ is fetched from the upstream server, stored in the cache, and transmitted to the user. 
PA-Cache manages a priority queue \textit{Q}, which stores the cached contents along with their predicted popularities. The head element of \textit{Q} is regarded as the least popular content which can be quickly retrieved under this data structure. Each eviction operation will update \textit{Q} accordingly. To keep the prediction of content popularity up-to-date, PA-Cache triggers a \textit{Learning Handler} periodically after every $\phi$ hours to update the prediction of content popularity. The PA-Cache algorithm is presented in Algorithm \ref{algorithm2}. 

\begin{algorithm}[t] 
	\caption{PA-Cache algorithm.}\label{algorithm2}
	\hspace*{0.02in} {\bf Input:}  Request $k$ 
	\begin{algorithmic}[1]
	 	\State Update features vector for $c_k$ in \textit{Feature Database};
	 	\If{$Z_{k+1}^{c_k} == 0$}
         	\State Remove the head element $c^{evict}$ from \textit{Q};
              \State Fetch $c_k$ from the upstream server;              
              \State Insert $c_k$ and its estimated popularity into \textit{Q};  
         \EndIf
          \State Serve the client with $c(k)$;
         \If{$t_k \  mod \  \phi == 0$}
                 \State Re-predict the popularity for all contents by Algorithm \ref{algorithm1};
                 \State Rebuild the priority queue \textit{Q};
        \EndIf
	\end{algorithmic}
\end{algorithm}

%% file: evaluation.tex
\section{Trace-driven Evaluation Results} 
\label{evaluation}








In this section, we conduct extensive evaluations with real-world traces to evaluate the performance of PA-Cache.

\subsection{Dataset}

The experiments are simulated on a real-world dataset derived from iQiYi\footnote{http://www.iqiyi.com}, which is the largest online VoD service provider in China.
This dataset contains 0.3 million unique videos watched by 2 million users over 2 weeks and has been widely used in previous works \cite{understanding-performance-of-edge,toward-smart-and-cooperative}.
The relevant information is recorded for each trace item as follows:
(i) The device identifier (anonymized), which is unique for each device and will be used to label different users.
(ii) Request time, which records the timestamp when the user request arrives; 
(iii) Video content, which involves the video name and some basic information, e.g., score, number of comments. In addition, we implement a crawler to collect more data to complement the features of videos (e.g., the area, type, language, length, publish date, director, performer).

\begin{table}[t]
\centering
\caption{Parameter values.}
\label{tab:parameter}
\begin{tabular}{|c|c|c|}
\hline
\textbf{Parameter} &\textbf{Value} & \textbf{Description} \\\hline
$C$ &10,000 &Number of contents \\\hline
$K$ &446629 &Number of requests \\\hline
$p$    &0.1\%-5.0\% &Cache percentage\\\hline
$L$    &10 &Maximum capacity of DNN\\\hline
$m$    &128 &Mini-batch size \\\hline
$\beta$    &0.99 &Discount factor \\\hline
$\kappa$    &100 &Smoothing parameter of ``noisy'' data \\\hline
$\zeta$ &0.1 &Smoothing parameter of minimum weight \\\hline
$\eta$ &10 &Learning rate \\\hline
$\phi$  &1h &Time window\\\hline
\end{tabular}
\end{table}

\subsection{Algorithm Implementation}

We implement a discrete event simulator based on the framework depicted in Fig. \ref{fig:structure}. \rb{To calculate and compare it with existing algorithms under restricted computational \rb{resources}, we randomly sample $C= 10,000$ videos from the dataset 9 times.} The cache percentage, which is the ratio between the cache size and the total number of unique contents, is considered to range from 0.1\% to 5.0\%. By default, it is set $p = 1.0\%$. The \textit{Learning Handler} module trains a DNN in Fig. \ref{fig:model} after every $1$ hour. 
The traces are divided into two periods. 
The first one is named \textit{warm-up} period. It spans the first seven days of request traces, representing the input of evolving DNN.
The second one is \textit{test} period, which begins after the \textit{warm-up} period.
Unless explicitly clarified, the experimental results presented are all obtained under the above settings. \rb{PA-Cache is run on a PC with an Intel(R) Core(TM) i5-7360U CPU @ 2.30GHz, NVIDIA GTX 1080 GPU, and 8 GB RAM using MXNet framework.} In this paper, we train a 10-layer evolving DNN with 512 units in the first layer and 16 units in the last hidden layer. The number of units from the $2^{nd}$ layer to the $9^{th}$ layer decreases gradually. 
Some key parameters, along with their descriptions and values, are listed in Table \ref{tab:parameter}.

\begin{figure*}[t]
\centering
\includegraphics[width=0.8\linewidth]{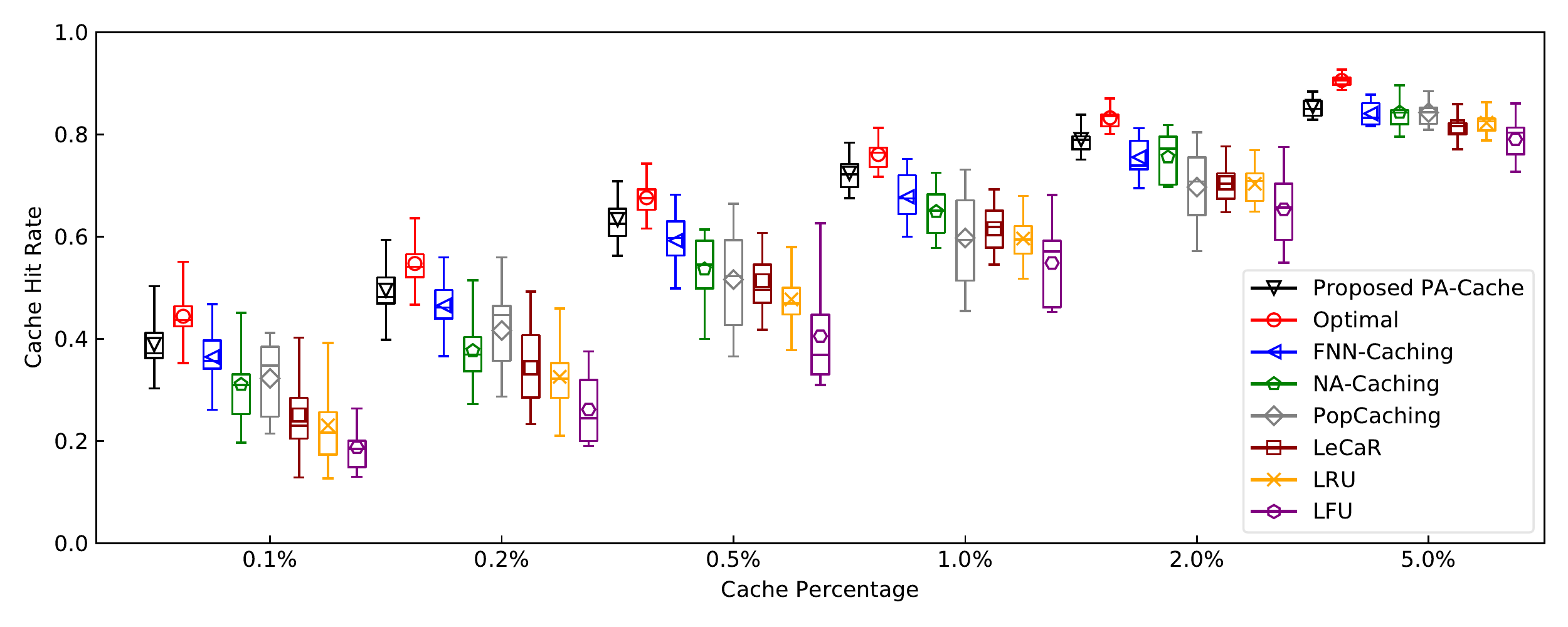}
\caption{\rb{Cache hit rate under different cache percentages.}}
\label{fig:hit-rate-with-cache-percentage}
\end{figure*}

\begin{figure*}[t]
\centering
\includegraphics[width=0.8\linewidth]{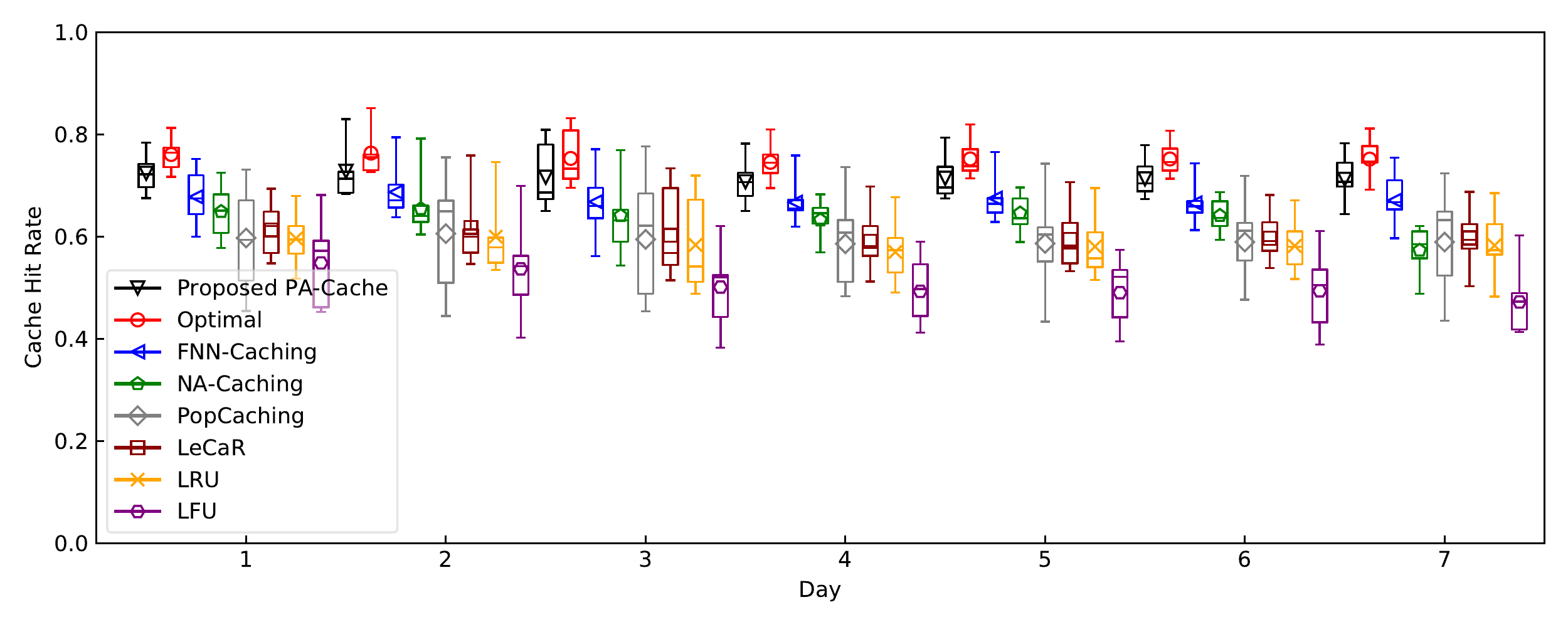}
\caption{\rb{Cache hit rate over time.}}
\label{fig:hit-rate-with-day}
\end{figure*}

\subsection{Benchmarks}
We compare our algorithm PA-Cache against the following baselines:

\begin{itemize}

\item \textbf{Optimal} \cite{a-study-of-replacement}. The cache runs an optimal and offline strategy for replacing the content, which has the longest time to be visited next time.

\item \textbf{FNN-Caching} \cite{feedforward-neural-networks-for}. The cache employs FNN to predict content popularity and accordingly makes caching decisions.

\item \textbf{NA-Caching} \cite{na-caching-an-adaptive}. The cache uses deep reinforcement learning from its own experiences to make caching decisions based on the features of dynamic requests and caching space.

\item \textbf{PopCaching} \cite{popularity-driven-content-caching}.  
The cache clusters different contents into hypercubes based on the similarity between their access patterns and predicts their popularity when making replacement decisions.

\item \textbf{LeCaR} \cite{driving-cache-replacement-with}. The cache maintains two history entries based on recency policy and frequency policy. The weight of each entry is updated by regret minimization. It is used to determine which policy to be applied for the cache eviction.

\item \textbf{LRU} \cite{analyzing-the-performance-of}. The cache manages an ordered queue which records the recent accesses of all the cached contents. When the cache is full, the least recently accessed content will be replaced by the \rb{newly} requested content.

\item \textbf{LFU} \cite{high-performance-cache-replacement}. The cache is implemented as a priority queue: the one that has been requested least frequently is replaced by the new content when the cache is full.

\end{itemize}

\subsection{Performance Comparison}

\begin{figure*}[t]
\centering
\includegraphics[width=0.8\linewidth]{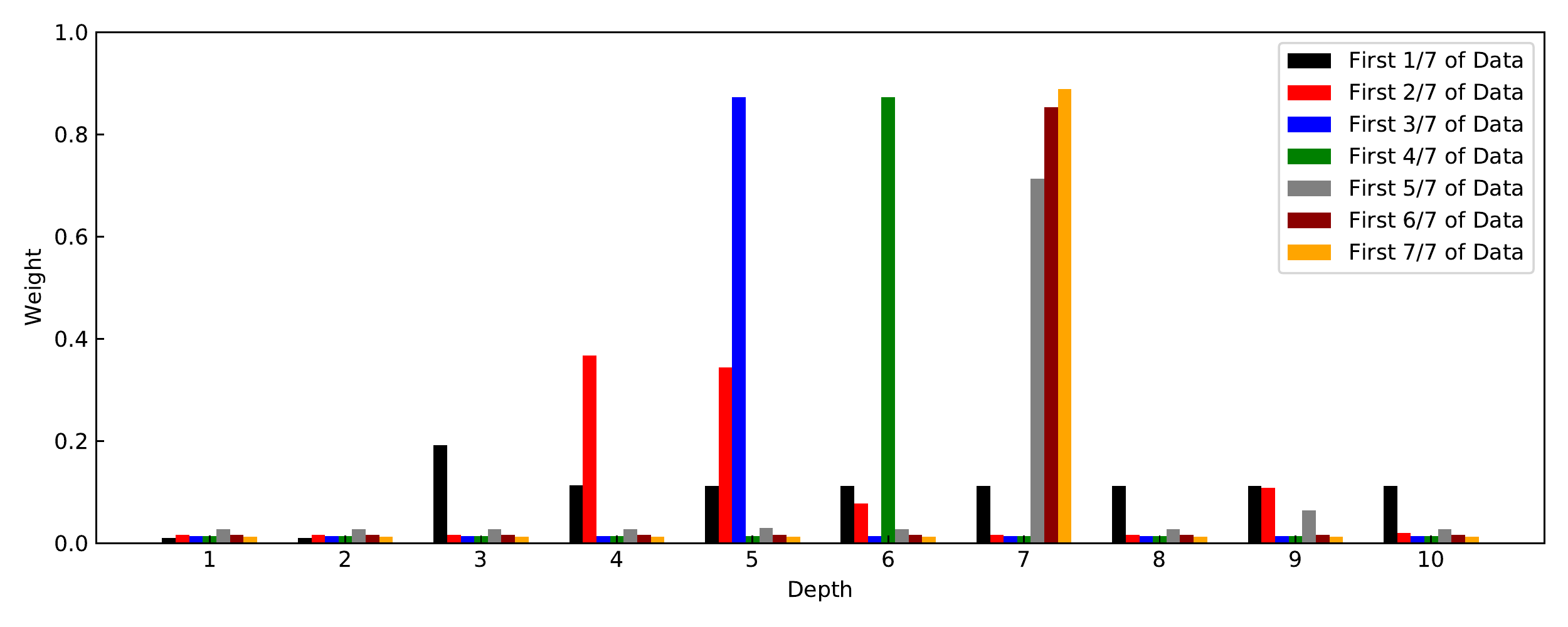}
\caption{\rb{Evolution of weight distribution of the DNN over time on the training set.}}
\label{fig:overall-weight-distribution}
\end{figure*}

\rb{The cache hit rate is depicted by the boxplot in Fig. \ref{fig:hit-rate-with-cache-percentage}, where the x-axis represents the cache percentage. The boxplot shows the means through the hollow markers, the medians through the bars, and the maximum and minimum values through the whiskers.
It illustrates that PA-Cache significantly outperforms LRU, LFU, and their combination LeCaR in all cases.  In particular, when the cache percentage is 0.1\%, PA-Cache's average performance advantage over LFU, LRU, and LeCaR exceed 107.1\%, 68.4\%, and 54.8\%, respectively. This is because these rule-based algorithms make eviction decisions based on simple metrics such as access time or frequency without considering the context information of contents. Moreover, when the content is evicted from the cache (e.g., LRU, LFU) or history (e.g., LeCaR), all information about it is missing and cannot be utilized for future decision making.}

\rb{Because the performance of the PopCaching algorithm is highly dependent on the hand-crafted features, we generate the content context vector several times and select the parameter setting, which achieves a $75^{th}$ percentile. It is observed that PA-Cache achieves a higher hit rate than FNN-Caching, NA-Caching, and PopCaching, especially when the cache percentage is small ($\leq 1.0\%$). For example, PA-Cache outperforms PopCaching, NA-Caching, and FNN-Caching by 22.8\%, 18.1\%, and 7.0\% in the case where cache percentage is 0.5\%. These results demonstrate that previous caching algorithms perform well under a certain scale of cache percentages, whereas PA-Cache can adapt to any cache percentages. For instance, when the cache percentage ranges from 0.2\% to 1.0\%, PA-Cache increases the hit rate by 7.9\%-12.7\% compared to PopCaching. Our interpretation is that PopCaching relies on the average sampling method for prediction, which is remarkable over hot contents but does not get an edge on lukewarm or cold contents.}

\rb{Fig. \ref{fig:hit-rate-with-day} depicts how the cache hit rate of the algorithms varies over time. On average, we can see that PA-Cache outperforms FNN-Caching, NA-Caching, PopCaching, LeCaR, LRU, LFU by 6.4\%, 13.2\%, 20.9\%, 19.0\%, 22.5\%, 42.1\%, respectively. 
PA-Cache approximates the Optimal with only a 3.8\% performance gap. This can be attributed to our model's prediction error.  Besides, For the average hit rate, PA-Cache maintains a more stable cache hit rate with a smaller variance of $4.2\times 10^{-5}$ compared to that of other methods (e.g., $5.0\times10^{-5}$ for FNN-Caching, $6.6\times10^{-4}$ for NA-Caching, $4.3\times10^{-5}$ for PopCaching, $8.7\times10^{-5}$ for LeCaR, $8.8\times10^{-5}$ for LRU, $6.4\times10^{-4}$ for LFU). In particular, NA-Caching and LFU exhibit the highest variances. This is because LFU predicts content popularity based on historical frequency without decay, which would be outdated over time.
NA-Caching suffers from large-delayed rewards for some contents, which might lead to convergence to poor action choices and slow policy learning.
On the contrary, PA-Cache could quickly adapt to changes in the workload over time.}

\rb{Fig. \ref{fig:overall-weight-distribution} illustrates the evolution of the weight distribution of the DNN over time on the training set for the sampled dataset, which achieves the median performance. 
Initially (first 1/7 data), the maximum weight has gone to the shallower regression in the $3^{rd}$ layer. Then (first 2/7 data), the maximum weights have moved to the regressions in the $4^{th}$ and $5^{th}$ layers. As more data arrives sequentially, deeper regressions will pick up higher weights. For instance, in the segment with the first 6/7 data, the $7^{th}$ layer has obtained the highest weight. It shows that our evolving DNN is capable of selecting an appropriate model automatically.}

\begin{figure}[t]
\centering
\includegraphics[width=0.85\linewidth]{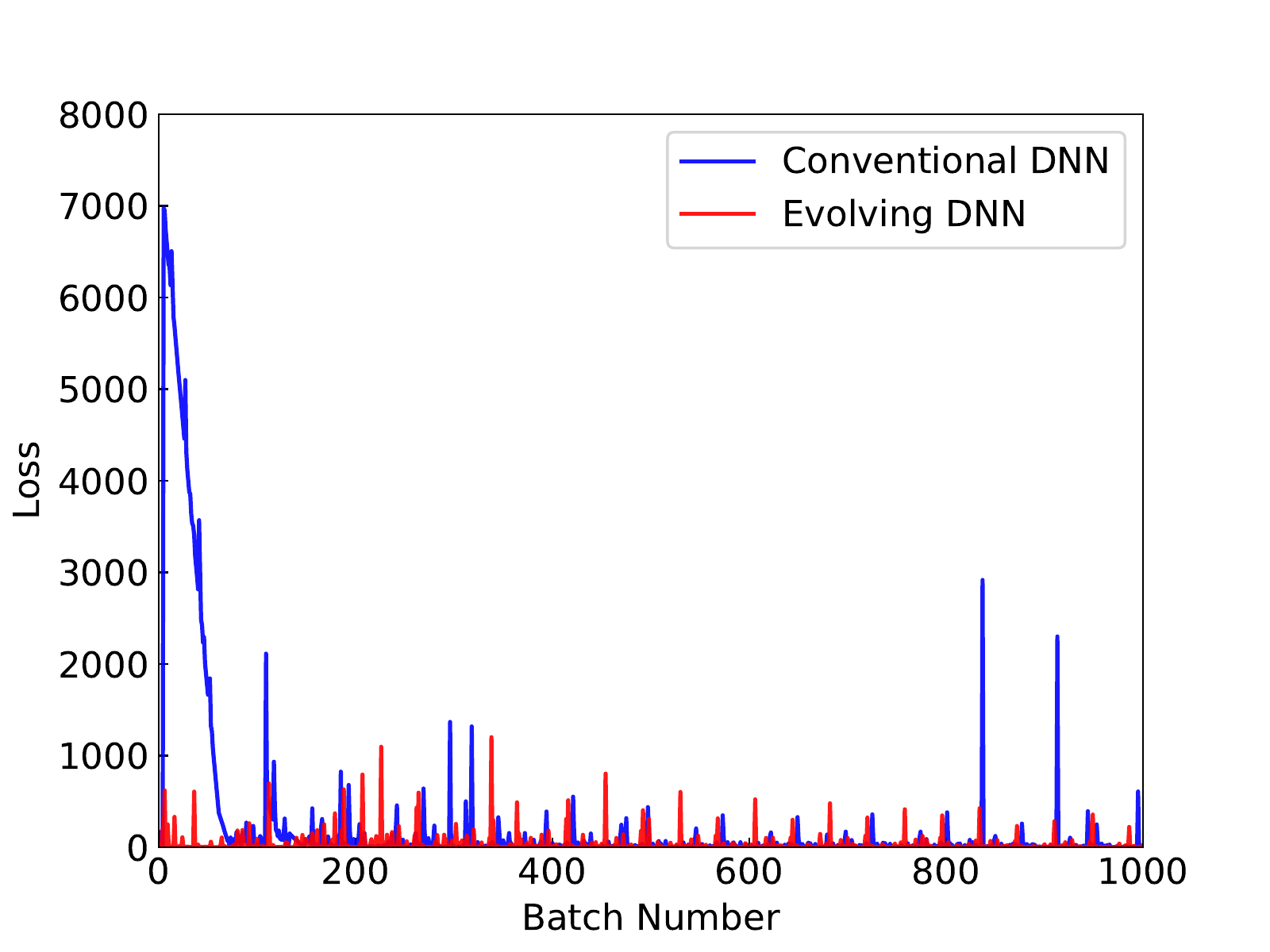}
\caption{Convergence behavior of the conventional DNN and our evolving DNN over batch number.}
\label{fig:batch-number}
\end{figure}

\begin{figure}[t]
\centering
\includegraphics[width=0.85\linewidth]{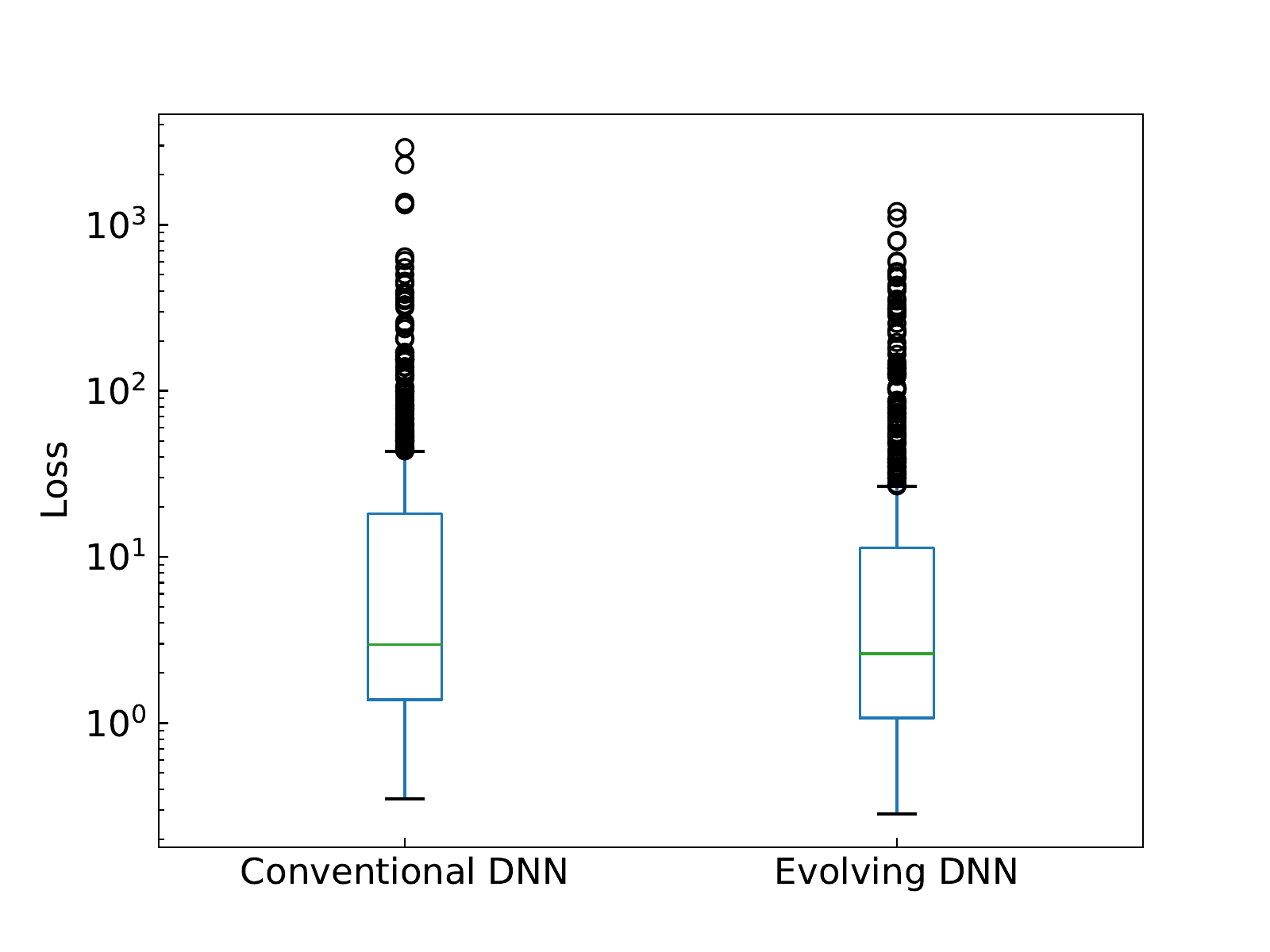}
\caption{Boxplots for 800 loss values during training after 200 batches by the conventional DNN and our evolving DNN.}
\label{fig:box-error}
\end{figure}

We compare the convergence behavior of our evolving DNN with a conventional 10-layer DNN in Fig. \ref{fig:batch-number}. It shows the variation of loss as the batch number increases. We can see that the loss of the 10-layer conventional DNN converges to a local optimum after about the $120^{th}$ batches, while our evolving DNN converges much more quickly. It means that our evolving DNN could benefit from the fast convergence of its shallow networks at the beginning. Moreover, Fig. \ref{fig:box-error} depicts the boxplots for 800 loss values obtained during the training after 200 batches by the conventional 10-layer DNN and our evolving DNN. 
In this plot, the boxes relate to the interquartile range; small circles denote the outliers; the upper and lower whiskers represent loss values outside the middle 50\%. We can observe that both the 25th and 75th percentile of evolving DNN are lower than those of conventional DNN. The median value of evolving DNN and conventional DNN is 2.97 and 2.61, respectively, \rb{represented by the bar} in the box. The boxplots indicate that the evolving DNN obtains both smoother variance and lower median loss. It illustrates that our evolving DNN keeps the merits of \rb{a powerful representation of the deeper network}.

\subsection{Practical Consideration}

\begin{figure}[t]
\centering
\includegraphics[width=0.85\linewidth]{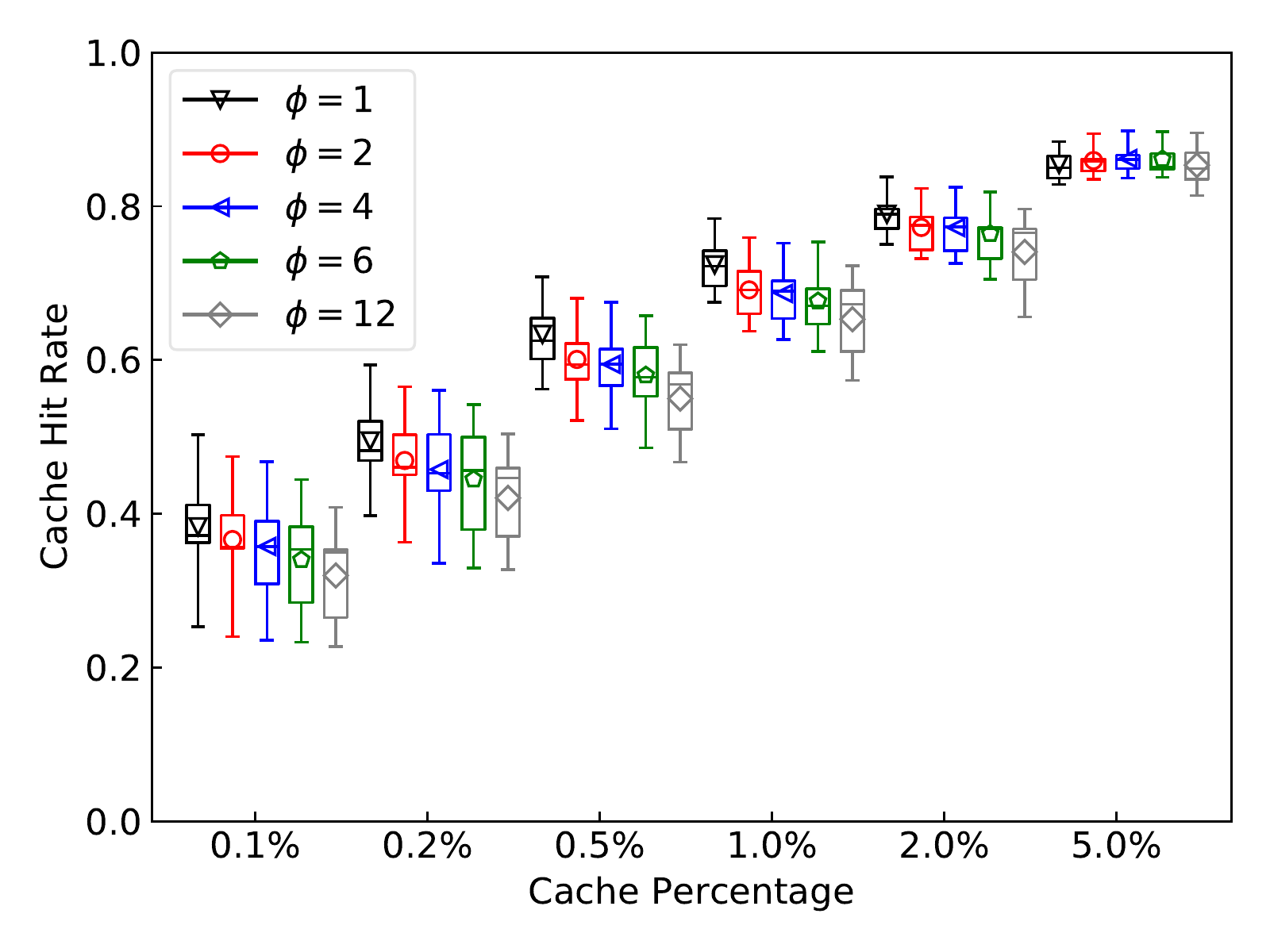}
\caption{\rb{Cache hit rate under different cache percentages for different $\phi$.}}
\label{fig:phi}
\end{figure}

\rb{The first consideration is the frequency of re-estimation for content popularity. 
Although the system is capable \rb{of correcting deviation between the predicted popularity and the real one} and adapt to time-varying demands more quickly if updating the evolving DNN more frequently, each update incurs additional overhead. Fig. \ref{fig:phi} illustrates how different values of parameter $\phi$ affect the caching performance under different cache percentages.
It is observed that the smaller the $\phi$ is, the higher the cache hit rate achieves In almost all cases, other than the case in which the cache percentage is 5.0\%. When the cache percentage ranges from 0.1\% to 2.0\%, the cache hit rate with $\phi=1$ is 2.2\%-5.4\% higher than that with $\phi=2$, but the performances under $\phi=2$ and $\phi=4$ are very similar.  After the cache percentage increases to $\phi  =5.0\%$, the cache hit rate is almost constant even if a different $\phi$ is selected. This observation indicates that the edge cache could select an appropriate parameter $\phi$ based on its storage capacity. That is, when the storage capacity is small, the cache node could optimize the parameter by seeking a trade-off between prediction accuracy and computational cost, whereas, for the large capacity, our algorithm scales to time windows with different sizes for updating.}

The historical information is available for existing contents, but new contents generated from content providers may be added to the system continually. Although the historical information of $\phi$ hours in our PA-Cache, much smaller than PopCaching or FNN-Caching, is enough for prediction. The cold start problem of new contents is still crucial for the performance improvement of \rb{the} caching system. We can employ a small LRU-cache space \cite{optimal-content-placement-for} to handle the traffic demand of new contents for which we do not estimate.

%% file: conclusion.tex
\section{Conclusion} 
\label{conclusion} 

In this paper, we have presented the design, implementation, and evaluation of PA-Cache, a novel popularity-aware edge content caching algorithm in edge networks. It makes adaptive caching decisions with the aim to maximize the cache hit rate in the long term. PA-Cache can effectively tackle the daunting task of content caching upon time-varying traffic demands. It has combined the strength of multi-layer RNN in learning the comprehensive representations of requested contents and the hedge backpropagation strategy that automatically determines how and when to accommodate network's depth in an evolving manner. The experiments on real-world traces have been conducted, and the performance of PA-Cache has been evaluated. Trace-driven evaluation results have demonstrated the effectiveness and superiority of our proposed PA-Cache in terms of the hit rate and computational cost compared to existing popular caching algorithms.